\documentclass[aps,prb,amsmath,amssymb]{revtex4-1}
\usepackage{graphicx}% Include figure files
\usepackage{epsfig,rotate,latexsym}
\usepackage{color, bm}
\usepackage{dashrule}
\begin{document}
\title {Electronic structure of Pr$_{2}$MnNiO$_{6}$ from x-ray photoemission, absorption and density functional theory}
\author {Padmanabhan Balasubramanian $^{1,2}$\footnote{bpaddy123@gmail.com}, Shalik Ram Joshi$^2$, Ruchika Yadav$^{3}$, Frank M. F. de Groot,$^{4}$, 
Amit Kumar Singh$^5$, Avijeet Ray$^1$, Mukul Gupta$^6$, Ankita Singh$^1$, Suja Elizabeth$^3$, Shikha Varma$^2$, Tulika Maitra$^1$, Vivek Malik $^1$}
\affiliation{$^1$Department of Physics, Indian Institute of technology, Roorkee-247667, Uttarakhand, India.\\
$^2$Institute of Physics, Bhubaneshwar-750012, India.\\  
$^3$Department of Physics,Indian Institute of Science, C.V. Raman Avenue, Bangalore-560012, India.\\
$^4$Inorganic Chemistry \& Catalysis, Debye Institute for Nanomaterials Science, Utrecht University, Universiteitsweg 99, Utrecht 3584 CG, The Netherlands.\\
$^5$Institute Instrumentation Centre, Indian Institute of technology, Roorkee-247667, Uttarakhand, India.\\
$^6$UGC-DAE Consortium for Scientific Research, University Campus, Khandwa Road, Indore 452 017, India.}
\date{\today}
\begin{abstract}
The electronic structure of double perovskite Pr$_{2}$MnNiO$_{6}$ is studied using core x-ray photoelectron spectroscopy and x-ray absorption spectroscopy.
The $2p$ x-ray absorption spectra show that Mn and Ni are in 2+ and 4+ states respectively. 
Using charge transfer multiplet analysis of Ni and Mn $2p$ XPS spectra,  we find charge transfer energies ${\Delta}$ of 3.5 and 2.5 eV for Ni and Mn respectively. 
The ground state of Ni$^{2+}$ and Mn$^{4+}$ reveal a higher $d$ electron count of 8.21 and 3.38 respectively as compared to the atomic values of
8.00 and 3.00 respectively thereby indicating the covalent nature of the system.
The O $1s$ edge absorption spectra reveal a band gap of 0.9 eV which is comparable to the value obtained from first principle calculations
for $U-J$ ${\geq}$ 2 eV.
The density of states clearly reveal a strong $p$-$d$ type charge transfer character of the system, with band gap proportional to average
charge transfer energy of Ni$^{2+}$ and Mn$^{4+}$ ions.
\end{abstract}
\maketitle
\section{Introduction}
\label{introduction} 
Transition metal compounds have always been of great interest since they show diverse physical properties like metal-insulator transition,
high temperature superconductivity, multiferroicity and various interesting phenomena like charge/orbital ordering and complex
magnetic structures\cite{Mott, Urushibara, Cava, Bednorz, Cheong, Taniguchi, Khomskii, Pavarini, Tokura, Lautenschlager, Tomioka, Li}. 
These systems include simpler oxides like NiO, MnO or more complex materials like KNiF$_{3}$, rare-earth manganites, cuprates
and nickelates\cite{FujimoriMnO, ShenNiO, RicartKNiF3, Tokura, Cava, Neidmayer1992}. 
In most cases, the parent compound is usually insulating, which becomes metallic under influence of doping or pressure\cite{Urushibara,Obradors}.
In a unified scenario, insulating behaviour of the various oxides (or sulphides, dihalides) 
can be described by the Zannen-Sawatzky-Allen(ZSA) phase diagram, which classifies the materials into Mott-Hubbard and charge-transfer insulators\cite{Sawatzky1985}. The electronic behaviour is governed mainly by three parameters namely Coulomb repulsion $U_{dd}$ in the $d$ orbital of the transition metal ion, ligand to metal charge transfer energy ${\Delta}$ and and metal-ligand hybridization strength $V_{pd}$\cite{Sawatzky1985}.
In the early transition metal (Ti, V and Cr) compounds, $U_{dd}$${<}$${\Delta}$ and the bandgap $E_{g}$${\propto}$$U_{dd}$. These materials are
known as Mott-Hubbard insulators. The late transition metal based compounds (eg: hole doped cuprates, NiCl$_{2}$, NiBr$_{2}$), show greater ligand-metal
charge transfer effect and for which $U_{dd}$${>}$${\Delta}$\cite{Ohta1991, Sawatzky1986}. Their band gap  $E_{g}$${\propto}$${\Delta}$, due to which
these materials are known as charge transfer insulators. 
In a charge transfer insulator, the ground state involves a strong fluctuation between $d^n$ and $d^{n+1}L$ states where $L$ is the ligand hole. However in
compounds involving Mn and Fe, the scenario is much more complicated and the band gap can be considered of an intermediate character.
In addition to ratios, $U_{dd}$/$V_{pd}$ and ${\Delta}$/$V_{pd}$, additional parameters like $3d$ bandwidth $W$ and anion bandwidth $w$ play an 
important role in determining whether the given compound is a metal or insulator.

 $2p$ x-ray photoemision(XPS) and x-ray absorption spectroscopy(XAS) are probes of $U_{dd}$, ${\Delta}$ and $V_{pd}$.   
Appearance of satellite peaks in the XPS spectra help in determining the three parameters using cluster analysis or single impurity Anderson model\cite{Bocquet1995}. 
The position and intensity of the satellite peaks systematically depend on the surrounding ligand. This is also an indicator of strength of covalency
and is related directly to the Slater-Koster transfer integrals $V_{pd{\sigma}}$ and $V_{pd{\pi}}$. The $2p$ XPS spectra of the late transition 
metal compounds is particularly sensitive to ${\Delta}$ and electronegativity of the anion \cite{Sawatzky1986}. However the XPS spectra is severely
broadened by multiplet and core hole effects. 
Complementary to XPS is the $2p$ XAS, which has certain advantages over XPS. Depending on the valency of the metal, the $2p$ XAS spectra has a distinct
shape. Also unlike XPS, which accesses the full multiplet, the no of transitions are restricted by the dipole transition rules. 

Core level XPS and XAS studies have been carried out on the rare-earth nickelates($R$NiO$_{3}$; $R$ = La, Nd, Pr..) and manganites ($R$MnO$_{3}$) in detail.
In the nickelates, studies have shown the variation of covalency and reduced hopping as we vary $R$ from La to Nd, causing changes
in conducting behaviour\cite{Medarde1992}. The Ni ion due to its high valence state of 3+ has a very small charge transfer
energy (${\Delta}$${\sim}$1 eV) leading to metallicity or an insulator with a very small band gap.
In $R$MnO$_{3}$ compounds, a larger band gap($E_{g}$${\sim}$1 eV) arises due to the Jahn-Teller effect at the Mn$^{3+}$ site,
In CaMnO$_{3}$ which is an Mn$^{4+}$ system though belonging to family of manganites, there occurs a large band gap due to the large 
crystal field splitting in Mn$^{4+}$ ion \cite{SinghLCMO1996}. 
In $R$MnO$_{3}$, the ground state shows a larger ${\%}$ of $d^{4}$ and a smaller ${\%}$ of $d^{5}L$ states with ${\Delta}$${\sim}$4-5 eV. However
in (Ca/Sr)MnO$_{3}$, with a smaller ${\Delta}$${\sim}$3 eV as seen from valence band and $2p$ core level photoemission is considered as a charge 
transfer insulator \cite{Saitoh1995}.

However, the homovalent substitution of Mn and Ni as in LaMn$_{1-x}$Ni$_{x}$O$_{3}$ leads to totally different ground state,
especially for x = 0.5. The half doped  compound, LaMn$_{0.5}$Ni$_{0.5}$O$_{3}$ also crystallizes as La$_{2}$MnNiO$_{6}$, depending on
synthesis technique\cite{Joly2002}. The former is orthorhombic($Pbnm$) while the latter belongs to the class of double perovskite 
compounds with monoclinic symmetry. In the orthorhombic structure, Mn and Ni ions are randomly arranged, since they occupy the
same Wycoff positon $2b$. However charge disproportionation results in formation of Mn$^{4+}$ and Ni$^{2+}$ by the following 
reaction, Ni$^{3+}$+Mn$^{3+}$ ${\rightleftharpoons}$ Ni$^{2+}$+Mn$^{4+}$\cite{Joly2002,Goodenough2003}. This favours a rocksalt
like arrangement of Mn and Ni resulting in monoclinic double perovskite compound La$_{2}$MnNiO$_{6}$. Our studies are based on 
the double perovskite material Pr$_{2}$MnNiO$_{6}$, which is relatively less explored. 
The parent compounds, PrMnO$_{3}$ and PrNiO$_{3}$ are A-type and G-type antiferromagnetic insulators respectively, while Pr$_{2}$MnNiO$_{6}$ is a
ferromagnetic insulator\cite{SinghPr2MnNiO62011}. The Mn$^{4+}$-Ni$^{2+}$ super-exchange interactions are ferromagnetic in nature, yield a transition
temperature, as high as 280 K in La$_{2}$MnNiO$_{6}$ \cite{RJBooth2009}. With decreasing cationic radii due to increasing $R$, the decrease 
in $<$Mn-O-Ni$>$ bond angle affects the exchange interaction and decreases the magnetic transition temperature. However even in perfectly 
ordered monoclinic structure, there occurs small percentage of randomness in distribution of Mn and Ni which are known as anti-site disorders. 
This result in Mn$^{4+}$-Mn$^{4+}$ and Ni$^{2+}$-Ni$^{2+}$ super-exchange interactions which are anti-ferromagnetic in nature. In the extreme 
limit of anti-site disorders and random occupancies, there occurs formation of Mn$^{3+}$ and Ni$^{3+}$ regions, which can result in
Mn$^{3+}$-Ni$^{3+}$-ferromagnetic super-exchange interactions. This results in second transition at lower temperature, sometimes leading 
to a glassy state at low temperatures\cite{Joly2002,Shi2011}. Using $3s$ XPS, one can probe the valence state of Mn with greater precision,
since the $3s$ splitting is proportional to the local spin of the Mn ion\cite{Galakhov2002}.

Irrespective of presence of anti-site disorders, even in the perfectly ordered Pr$_{2}$MnNiO$_{6}$, the resultant local electronic structure is
different from both parent compounds. The combined overlap of Mn-O and Ni-O orbitals would affect the values of ${\Delta}$, $U_{dd}$ and $V_{pd}$. 
Thus it would be interesting to obtain an estimate of these parameters which lead to the ferromagnetic super-exchange interaction and also probe the
conducting behaviour(if metallic) or nature of the band gap as per the classification in the ZSA phase diagram. 
In the present paper we have used XPS and XAS accompanied by cluster-model calculations and density functional theory methods, 
to investigate the electronic structure of double perovskite Pr$_{2}$MnNiO$_{6}$. In addition, the O $1s$ edge XAS spectra was also
obtained and compared with unoccupied density of states along with estimation of the band gap.

\section{Methodology}
\subsection{Experimental}
The polycrystalline samples of Pr$_{2}$NiMnO$_{6}$ a were synthesized by conventional solid-state reaction. Resistivity studies were carried out in
temperature range of 4 to 300 K using four probe method. Magnetic properties were measured using a superconducting quantum interference device (SQUID)
in the temperature range 10 - 300~K. AC susceptibility measurements were carried out in a commercial CYROBIND set-up in the temperature range 4.2 - 280~K.
XPS studies were carried out using Al K${\alpha}$ source with a hemisphere analyzer with a resolution of 0.5 eV. The binding energies were calibrated
w.r.t C $1s$ photoelectron line with binding energy of 284.6 eV. The spectra was collected at the Mn and Ni ${\it 2p}$ edges, Mn ${\it 3s}$ edge along 
with O $1s$ edge. The XAS studies were carried out at BL-01 beamline in INDUS synchrotron centre, India at room temperature. The XAS spectra was obtained 
at O $1s$ edge, Mn and Ni $2p$ edges using the total electron yield method. The resolution of spectra was 0.1 eV.

%================================================================================================
%Computational
%================================================================================================
\subsection{Computational Studies}
Computational studies were performed using the projector augmented wavefunction (PAW) method within the density functional theory.
The ab-initio simuation package (VASP) was used for this purpose\cite{kresse1996}. The Pr $5d$, Mn $3d/4s$, Ni $3d/4s$, O $2s$ O $2p$ are
considered as valence orbitals while the Pr $4f$ orbitals are considered as core levels. The calculations were performed within the generalized 
gradient approximation(GGA) formalism. Both GGA and GGA+U formalism was used to see the effect of Coulomb correlations\cite{Anisimov1993}. 
The plane wave basis was used with a cutoff of 600 eV.  
Initially the crystal structure was relaxed until the forces on the atoms are less than 0.05 eV/{\AA} . The structural 
optimisation was carried out assuming a ferromagnetic ordering between the Mn and Ni spins in accordance with experiments. 
Then the selfconsistent electronic calculations were performed till the energy difference between successsive cycles were less
than 10$^{-5}$ eV. The band structure was obtained along specific directions along the Brillouin zone, while the partial spin
polarized density of states were obtained by performing integrations using a 7 x7 x5 Monkhorst pack. 
 
\section{Results and discussion}
The x-ray powder diffraction data of Pr$_{2}$MnNiO$_{6}$ was refined to monoclinic space group $P2_1/n$. The structural parameters
obtained from our refinement are $a$ = 5.4672{\AA} $b$ = 5.5362{\AA} and $c$ =  7.7336{\AA} with ${\beta}$ = 89.88$^{\circ}$. 
The average Mn-O and Ni-O bondlengths are 1.93 and 2.04{\AA} respectively. The parameters are well in agreement with the reported 
values\cite{RJBooth2009}. The three distinct Mn-O and Ni-O bond lengths in the octahedra are almost equal indicating absence of any 
disortions. The bond valence sums of 3.13 and 2.2 are nearly equal to the valencies expected from Mn$^{4+}$ and Ni$^{2+}$ system.
\begin{figure}[!h]
\begin{center}
   	\includegraphics[width=0.8\textwidth]{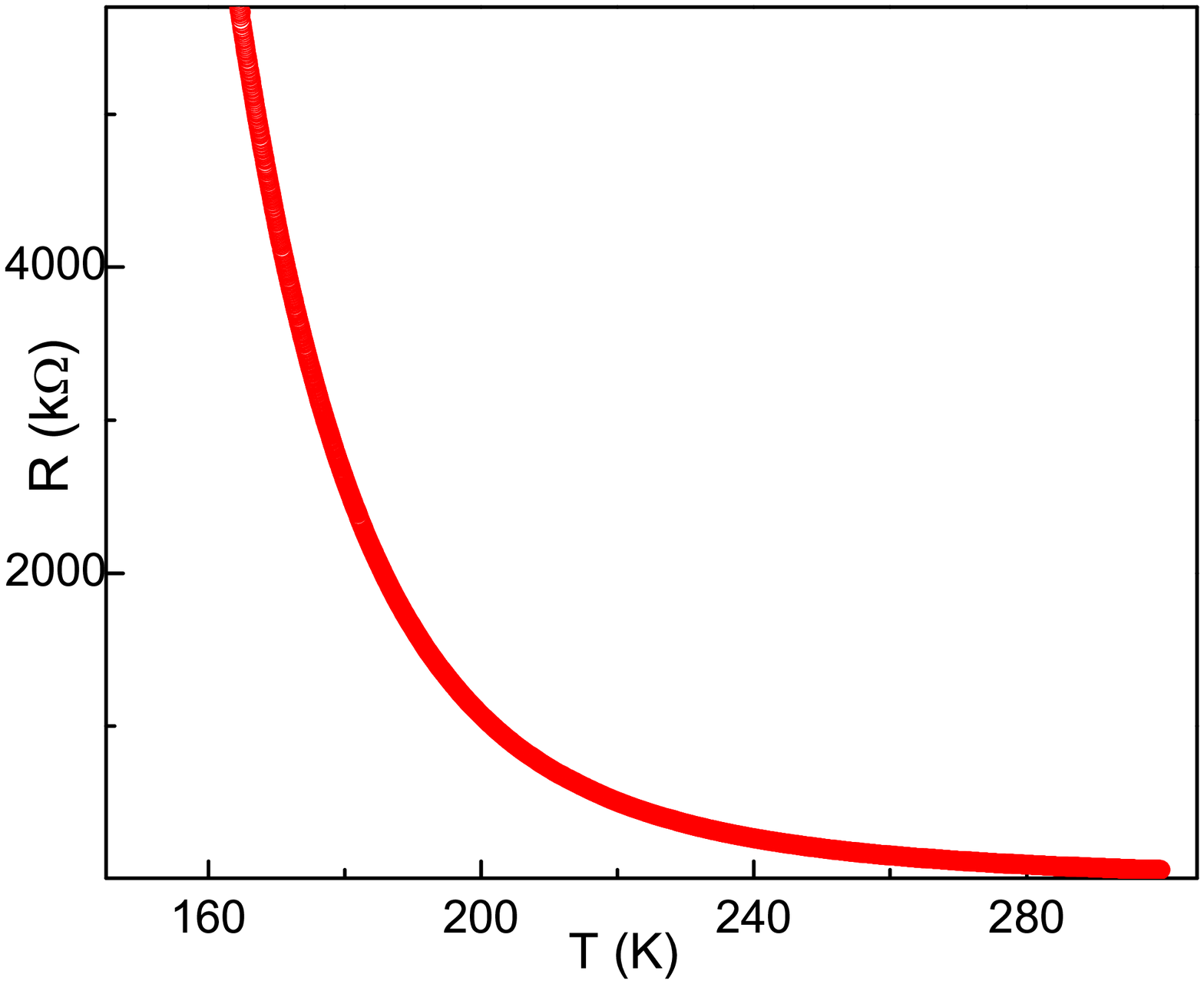}
	\caption{\textit{Temperature dependenence of resisitance of Pr$_{2}$NiMnO$_{6}$.}}
\end{center}
\end{figure}
Fig.1 shows plot of resistivity vs. temperature of Pr$_{2}$MnNiO$_{6}$. The plot is shown only till 160 K, since the value of
resisitivity becomes several mega-ohms below this temperature. Our material is insulating with an activation energy of 0.3 eV. 
The smaller value of activation energy is in agreement with the predicted band gap as obtained from GGA-based calculations as we 
would discuss below.
\begin{figure}[!h]
\begin{center}
   	\includegraphics[width=0.8\textwidth]{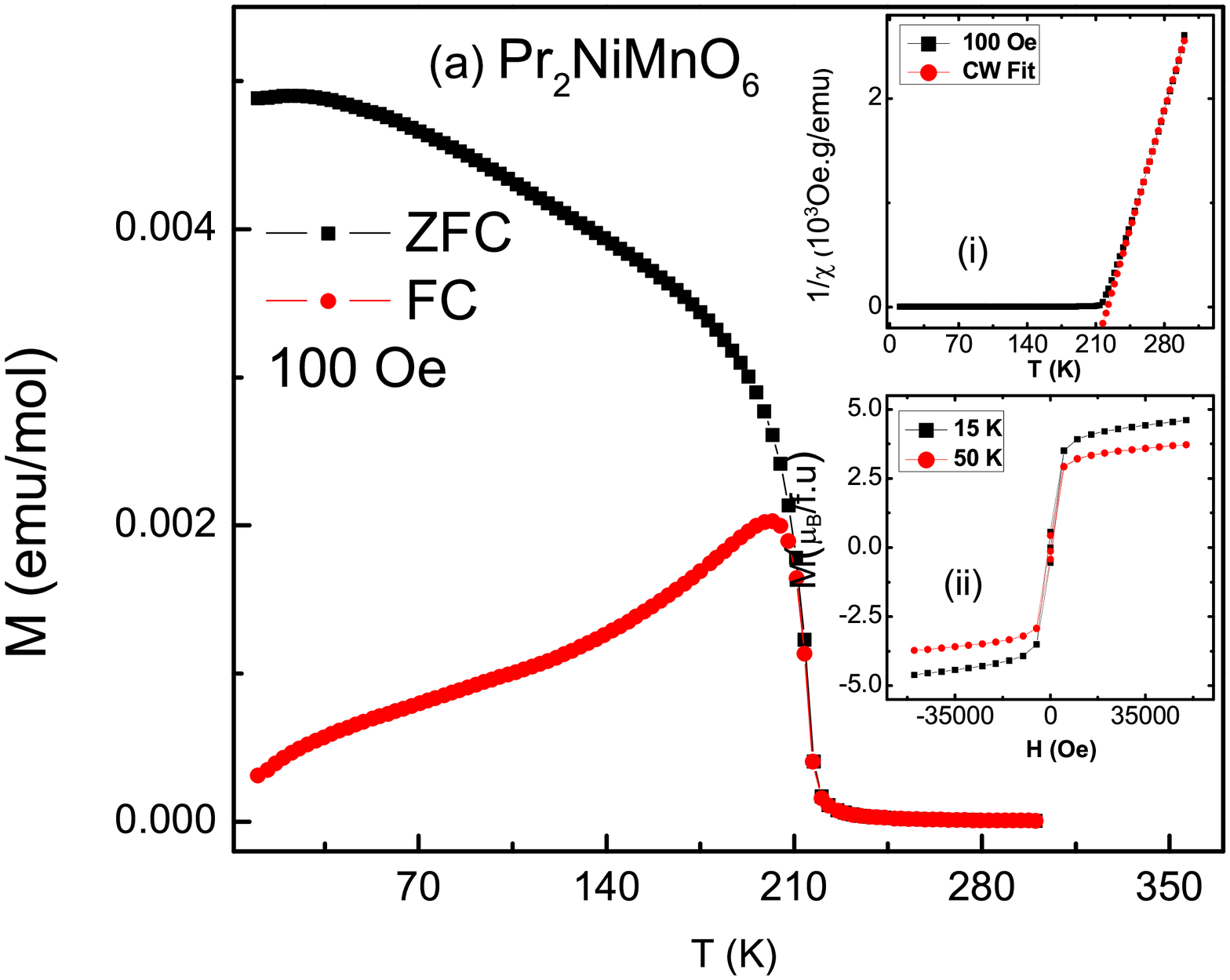}
   	\caption{\textit{ ZFC-FC magnetization of Pr$_{2}$MnNiO$_{6}$ in a field of 100 Oe. 
The top inset shows inverse susceptibility vs temperature and the corresponding Curie-Weiss fitting. The lower inset shows the M vs H at 15 K and 50 K.}}
\end{center}
\end{figure}
Fig.2 shows magnetization plots of Pr$_{2}$MnNiO$_{6}$ for ZFC and FC cooling. The paramagnetic-ferromagnetic transition occurs 
at 210 K which arises due to O$^{2-}$ mediated Mn$^{4+}$-Ni$^{2+}$ super-exchange interactions. 
%
%Existence of anti-site disorder induces a second magnetic transition at lower temperatures in many of the studied double perovskites. 
Absence of second transition in our system indicates a very low concentration of anti-site disorders.The magnetic moment
of Pr$_{2}$MnNiO$_{6}$ at 15 K and 5 T is around 4.8 ${\mu}_{B}$ which is close to the expected value of 5 ${\mu}_{B}$ due 
to perfectly ordered system. However the slightly reduced value and absence of complete saturation in magnetization suggest 
presence of small amount of anti-site disorders, in addition to role of Pr$^{3+}$ spins. 

\subsection{XPS and XAS spectra}
\subsubsection{Mn $3s$ spectra}
\begin{figure}[!h]
\begin{center}
   	\includegraphics[height=0.6\textwidth]{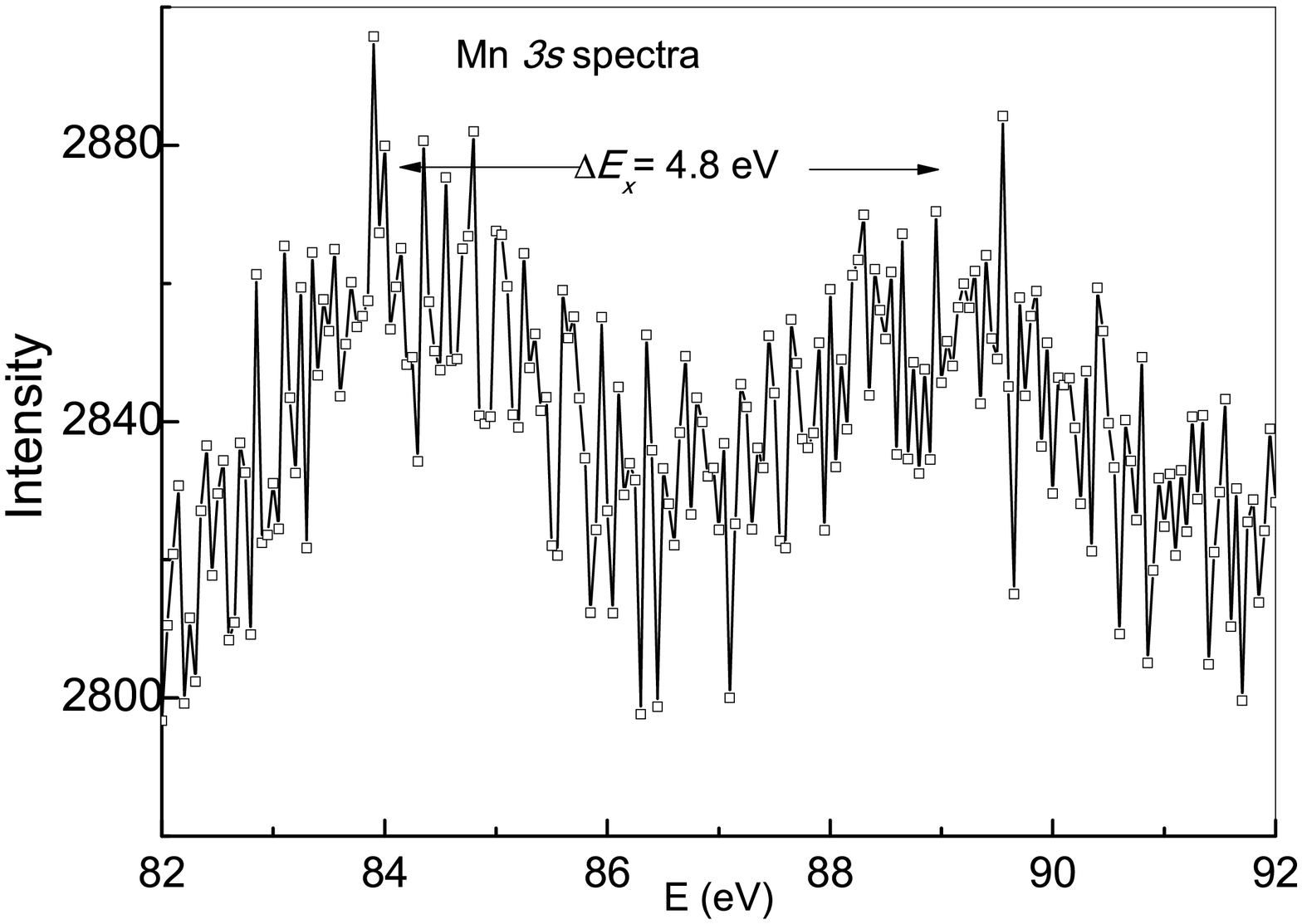}
	\caption{\textit{Mn ${\it 3s}$ XPS spectrum of Pr$_{2}$NiMnO$_{6}$ where the two peaks indicate exchange splitting.}}
\end{center}
\end{figure}
The role of anti-site disorders in affecting the total magnetic moment is obtained more precisely from Mn $3s$ XPS spectra.
 The $3s$ spectra arises due to transition from the initial $|$$3s$$^{2}$, $3d$$^{3}$$>$ to $|$$3s$$^{1}$, $3d$$^{3}$$>$ states.
The Mn $3s$ spectra shown in fig.3 exhibits a characteristic doublet due to the exchange splitting. The difference between interaction 
of $3s$ electron with the parallel and anti-parallel spin states of the $3d^n$ shell gives rise to the exchange splitting. 
This scenario is valid only in the case of early transition metal ions. In the case of Ni $3s$, the effect of charge transfer
reduces the observed exchange splitting.
The magnitude of splitting is proportional the Slater exchange integral $G^{2}$($3s$,$3d$) which is given by Van-Vleck theorem as\cite{VanVleck1934},
\begin{equation}
{\Delta E_{ex} =  \left(\frac{(2S+1)G^2(3s,3d)}{2l+1}\right)},              
\end{equation}
where $l$=2. The magnitude of ${\Delta}E_{ex}$ increases with decreasing valency\cite{Galakhov2002}. 
Our experimental spectra reveals an exchange splitting of nearly 4.8 eV. Assuming that $G^2(3s,3d)/(2l+1)$ = 1.1 eV, 
the above equation yields a net spin of $S$=1.68\cite{Manella2008}. For a complete Mn$^{4+}$ system like CaMnO$_{3}$, 
the value of ${\Delta}E_{ex}$ is nearly 4.5 eV\cite{Galakhov2002}, which yields a spin of $S$ = 3/2.
Thus in our material the value of $S$ is very close to the expected spin value of $S=3/2$ in an Mn$^{4+}$ system. 

The slightly higher value of $S$, indicates presence of Mn$^{3+}$ ions which arises due to anti-site disorders 
and mixed occupancy of the Ni(Mn) sites. However the signature of Mn$^{3+}$ is not so evident from our Mn $2p$ XPS and XAS spectra. 

\subsubsection{Simulation of Ni and Mn $2p$ XPS and XAS spectra}
In this section,we briefly discuss the theoretical simulation of the $2p$ XPS and XAS spectra.
The simulations were done in the configuration interaction cluster model, using charge transfer multiplet program CTM4XAS\cite{deGroot2005}. 
The simulations were performed for a single ion of Ni$^{2+}$ and Mn$^{4+}$ surrounded by oxygen ligand octahedra in $O_{h}$ symmetry. 
The ground state electron configuration of Ni$^{2+}$ is $d^{8}$, which in $O_{h}$ symmetry can be written as $^{3}A_{2g}$($t_{2g}^{6}$$e_{g}^{2}$).
We consider two charge transfer configurations, $d^{9}L$ and $d^{10}L^2$, where $L$ corresponds to a ligand hole in the O $2p$ state.
The Mn$^{4+}$ ion has $d^3$ configuration in ground state, which can be written as $^{4}A_{2g}$($t_{2g}^{3}$) in $O_{h}$ symmetry. 
The calculation of Mn $2p$ spectra involving two charge transfer configurations is computationally difficult. Hence we consider
only the $d^{4}L$ configuration. Also, effects of the $d^{5}L^2$ configuration is not so prominent in the $2p$ XPS spectra. 
The ground state wavefunctions for Ni$^{2+}$ and Mn$^{4+}$ ions are given as,
\begin{equation}
%\label{eqn1}
{\Psi}^{Ni}_{g}= \alpha_{0}|d^{8}>+\beta_{0}|d^{9}L>+\gamma_{0}|d^{10}L^2> .
\end{equation}
\begin{equation}
%\label{eqn1}
{\Psi}^{Mn}_{g}= \alpha_{0}|d^{3}>+\beta_{0}|d^{4}L> .
\end{equation}
 The ligand-metal charge transfer energy is defined as, ${\Delta}$=E($d^{n+1}L$)-E($d^{n}$)($<$$d^{n}$$|$H$|$$d^{n}$$>$-$<$$d^{n+1}L$$|$H$|$$d^{n+1}L$$>$),
 where $H$ is the model Hamiltionian describing the ground and excited states as mentioned by Okada {\it et al.} \cite{Okada1992}. The $d^{n+2}L^2$ state 
 occurs at a much higher energy, given by E($d^{n+2}L^2$)-E($d^{n}$)=2${\Delta}$+$U_{dd}$, where $U_{dd}$ = E($d^{n-1}$)+E($d^{n+1}$)-2E($d^{n}$) is 
 the $d$-$d$ Coulomb interaction\cite{Okada1992}. 
 The off-diagonal matrix elements, $V$ = $<$$d^{n}$$|$H$|$$d^{n+1}L$$>$ = $<$$d^{n}L$$|$H$|$$d^{n+2}L^2$$>$ which are the one-electron
 transfer integrals correspond to the metal-ligand hybridization. 
The anisotropy in $V$ due to splitting between the $t_{2g}$ and $e_{g}$ states in Ni$^{2+}$ and Mn$^{4+}$ are denoted as $V_{e_{g}}$ and $V_{t_{2g}}$ respectively.
In our calculations, $V_{t_{2g}}$ is fixed at 1 eV, thus $V_{e_{g}}$ is the single adjustable parameter. 
The hybridizations strengths are related to Slater-Koster transfer integrals, through the expressions
$V_{e_{g}}$ = -${\sqrt{3}}$$V_{pd\sigma}$ and $V_{t_{2g}}$ =-2$V_{pd\pi}$\cite{Bocquet1992}. 
The final state involves effect of $2p$ core hole which reduce final state energies by a constant term. This 
term, $U_{dc}$ corresponds to the attractive potential between the $2p$ core hole and the $3d$ electron. In the case of $2p$ XPS spectra the final states are,
\begin{equation}
%\label{eqn1}
{\Psi}^{Ni}_{f}= \alpha|\underline{c}d^{8}>+\beta|\underline{c}d^{9}L>+\gamma|\underline{c}d^{10}L^2> .
\end{equation}
\begin{equation}
%\label{eqn1}
{\Psi}^{Mn}_{f}= \alpha|\underline{c}d^{3}>+\beta|\underline{c}d^{4}L> .
\end{equation}
In the case of $2p$ XAS, the final states which are of the type $2p^{5}3d^{n+1}$ are given by,
\begin{equation}
%\label{eqn1}
{\Psi}^{Ni}_{f}= \alpha_{1}|\underline{c}d^{9}>+\beta_{1}|\underline{c}d^{10}L>.
\end{equation}
\begin{equation}
%\label{eqn1}
{\Psi}^{Mn}_{f}= \alpha_{1}|\underline{c}d^{4}>+\beta_{1}|\underline{c}d^{5}L> .
\end{equation}
In the above equations \underline{$c$} denotes the core-hole wavefunction.
The calculations were performed for the entire multiplet spectrum\cite{Okada1992}. The $3d$$-$$3d$ and $2p$$-$$3d$ Slater
integrals were reduced to 80\% of the Hartree-Fock values. 
 The effect of bare crystal field splitting between the $t_{2g}$ and $e_{g}$ states were also included in the calculation 
 by varying the separation 10$Dq$  between 0 to 2.5 eV. 
The intensity of XPS and XAS spectra are calculated using sudden approximation\cite{Sawatzky1986, Okada1992}. For matching 
the calculated spectra with experiments, the values ${\Delta}$, $V$ and $U_{dd}$ were systematically varied. Similarly, the
ratio $U_{dd}$/$U_{dc}$ was varied between 0.8 and 0.9 for optimum matching between experimental and calculated spectra.
The defenition of ${\Delta}$ and $U_{dd}$ are based on the centre of gravity of multiplet of each charge transfer configuration.
However, their actual values are defined based on the difference between the lowest multiplet energy of each configuration.
These values are appropriately labelled as ${\Delta}_{eff}$ and $U_{eff}$. Both the parameters, play a major role in determining
the ground state electronic properties of the system.
\subsubsection{Ni $2p$ XPS spectra}
\begin{figure}[!h]
\begin{center}
   	\includegraphics[height=0.6\textwidth]{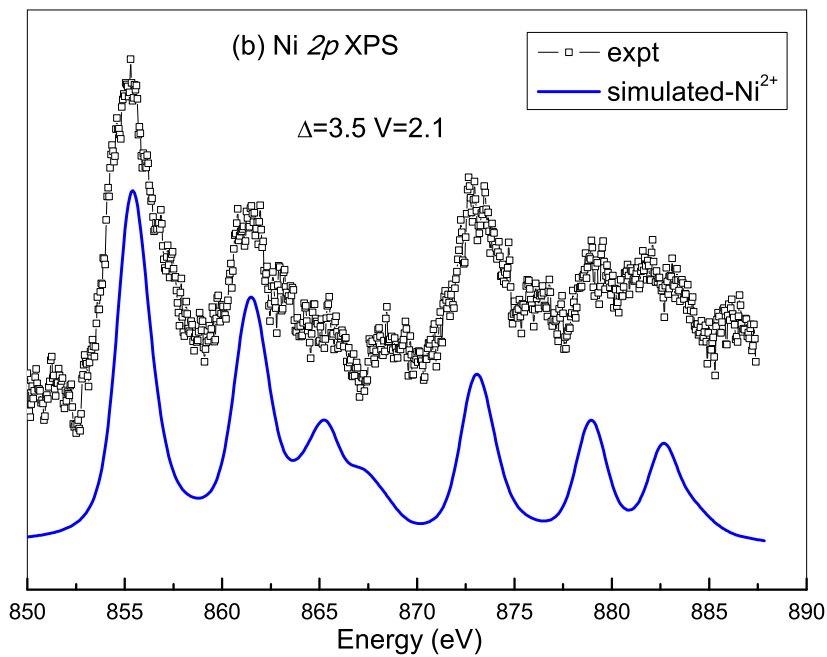}
	\caption{\textit{Ni ${\it 2p}$ XPS spectrum of Pr$_{2}$NiMnO$_{6}$ along with simulated spectra. The scales have been shifted for clarity.}}
\end{center}
\end{figure}

In Fig. 4, we show the Ni $2p$ XPS spectra of Pr$_{2}$MnNiO$_{6}$. The spectra shows spin-orbit split $2p_{3/2}$ and $2p_{1/2}$ 
regions with the peaks located at 855 and 875 eV, respectively. Both the $2p$ regions contains two additional satellite features 
in addition to the main peak. The second satellite of the $2p_{3/2}$ is less prominent and has a lower intensity as compared to
both the satellite peaks of $2p_{1/2}$. The second satellite of $2p_{1/2}$ is considerably broadened. 
The Ni$^{2+}$ ion in Pr$_{2}$MnNiO$_{6}$ is surrounded by oxygen octahedra, similar to that in NiO. However the XPS spectra of 
NiO shows only a single satellite peak at each edge, which is seen in the bulk as well as thin films of NiO \cite{Bocquet1995,Sawatzky1986,SawatzkyfilmsNiO1996}. 
 The second satellite feature is prominently seen nickel dihalides, NiCl$_{2}$ and NiBr$_{2}$ while NiF$_{2}$ shows only 
 a single satellite peak\cite{Sawatzky1986,Okada1991}. Among the Ni dihalides, NiF$_{2}$ has the largest value of ${\Delta}$ due 
 to the high electronegativity of fluorine, while ${\Delta}$ is smallest in the case of NiI$_{2}$. The prominent appearance of
 the second satellite clearly indicates a lower charge transfer energy and a greater covalency in our material as compared to
 the highly ionic character of NiO and NiF$_{2}$.  

The simulated Ni $2p$ XPS spectra for an NiO$_{6}$ cluster is shown in fig. 4. The spectra is 
optimised for ${\Delta}$ = 3.5 eV and hybridization ratio $V_{e_g}$/$V_{t_{2g}}$ = 2.1 eV to match with the experimental spectrum. 
The values of $U_{dd}$ and $U_{dc}$ are 7.5 and 9.0 eV respectively. The higher values of $U_{dd}$ are in agreement with 
the greater charge transfer character of the late transition metal ion compounds.
From the relative values of ${\Delta}$, $U_{dd}$ and $U_{dc}$, the ground states and final states of the XPS spectra can be 
classified in four regimes \cite{Sawatzky1986}. 
In our material since, ${\Delta}$ $>$ 0, the ground state has the following energy level sequence, E($d^{8}$)$<$ E($d^{9}L$)$<$ E($d^{10}L^2$). 
As the three  parameters satisfy the following inequalities, 2${\Delta}$+$U_{dd}$ $<$ $U_{dc}$ and $U_{dc}$ $-$ $U_{dd}$ $<$ ${\Delta}$ $<$ $U_{dc}$ $-$ $U_{dd}$/2,
in the final state, the level energy level sequence becomes E(c$d^{9}L$) $<$ E(c$d^{10}L^2$) $<$ E(c$d^{8}$). Thus in fig. 4, the main peak
has a majority $cd^{9}L$ character while the first and second satellite peaks have majority $cd^{10}L^2$ and $cd^8$ characters.
The position and intensity of the satellites are dependent on the ratio of ${\Delta}$/$V$. The weights of the $d^{8}$, $d^{9}L$ and $d^{10}L^2$ components
in the ground state are 0.78701, 0.20697 and 0.00603 respectively, which yields the average electron number $<$$n_{d}$$>$ = 8.21 in the ground state. 
 Using the relation mentioned by Fujimori \emph{et al.} \cite{Fujimori1993}, we have determined ${\Delta}_{eff}$ and $U_{eff}$, which
 are mentioned in table I. We find that ${\Delta}_{eff}$$>$${\Delta}$ and $U_{eff}$$<$$U$ as seen in the late transition metal ions. 
 Thus analysis of Ni $2p$ XPS spectra suggests that the Ni-O bond in Pr$_{2}$MnNiO$_{6}$ has an intermediate covalent character.
\subsubsection{Ni $2p$ XAS}
\begin{figure}[!h]
\begin{center}
   	\includegraphics[height=0.6\textwidth]{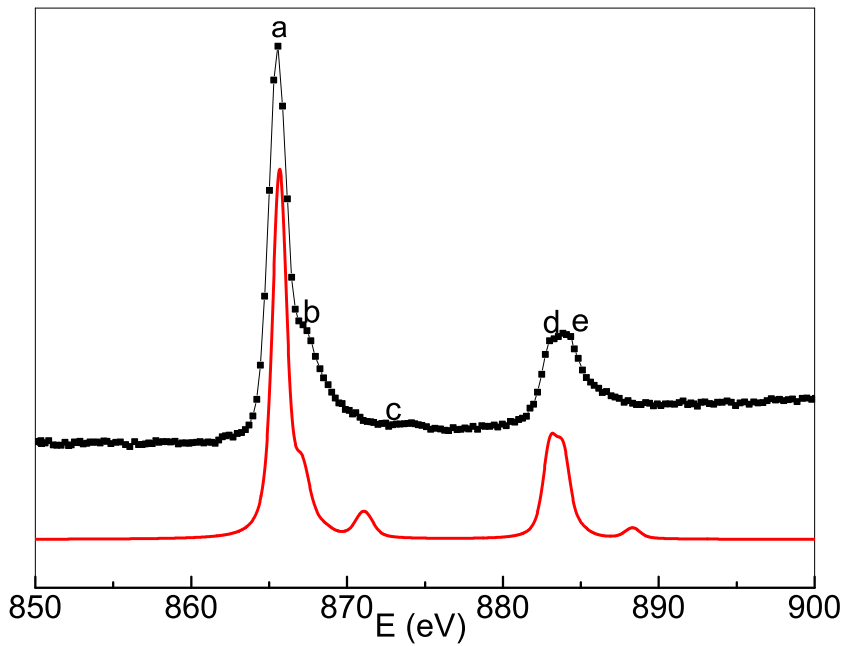}
	\caption{\textit{Ni $2p$ absorption spectra of Pr$_{2}$NiMnO$_{6}$ along with simulated spectra.}}
\end{center}
\end{figure}
The Ni $2p$ absorption spectra shown in Fig. 5, is split due to spin-orbit coupling into $2p_{3/2}$ and $2p_{1/2}$ peaks. The spectra displays 
characteristic feature of an Ni$^{2+}$ system\cite{Sawatzkyb1986}. Compared to the XPS spectra, the satellite intensities are weaker. 
The absorption spectra contains distinct features marked $a$ to $f$ as shown in fig. 5.  
The separation between the main peak $a$ and the shoulder peaks $b$ and $c$ and also the shape of the peaks are affected by ${\Delta}$ and $V$.
The XAS spectra qualitatively resembles the spectra of NiBr$_{2}$ and NiCl$_{2}$\cite{Sawatzkyb1986}.
In fig. 5, we also show the simulated Ni$^{2+}$ XAS spectra, which was obtained for ${\Delta}$ = 3.5 eV and $V$ = 2.1 eV. Unlike the XPS spectra, 
we have assumed only a single charge transfer configuration $d^{9}L$ in the ground state since the $3d$ states becomes filled for $d^{10}L^2$ configuration.
The spectra was broadened by convoluting the line spectra with a Lorentzian function (0.3 eV) and Gaussian function (0.4 eV). 
The spectra also shows an additional broadening, especially for features $b$ and $c$ compared to that observed in the Ni dihalides. This can
be attributed to the effect of presence of Ni$^{3+}$ ions in the system due to random occupancies by Mn/Ni. 
\subsubsection{Mn $2p$ XPS spectra}
\begin{figure}[!h]
\begin{center}
   	\includegraphics[height=0.6\textwidth]{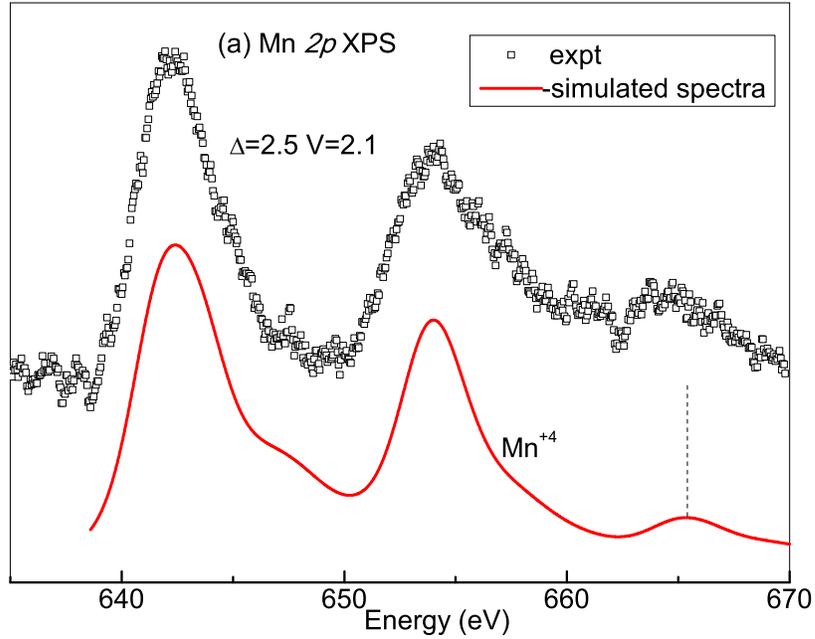}
	\caption{\textit{Mn ${\it 2p}$ XPS spectrum of Pr$_{2}$NiMnO$_{6}$ along with simulated spectra.}}
\end{center}
\end{figure}
In Fig. 6, we show the Mn $2p$ XPS spectra of Pr$_{2}$MnNiO$_{6}$. The spectra shows $2p_{3/2}$ and $2p_{1/2}$ spin-obit doublet
peaks located at 642 and 654 eV, respectively. In addition we observe the satellite peak of $2p_{1/2}$, at a binding energy of 666 eV. The satellite peak of $2p_{3/2}$ is not visible since it overlaps with the $2p_{1/2}$ peak. The position of the satellite peak w.r.t the main peak is sensitive to the $d$-electron count\cite{Abbate2002}. In fig.6 we also show the calculated Mn $2p$ spectra for an MnO$_{6}$ cluster. The calculated spectra is broadend with a energy dependent Lorentzian and Gaussian function of 0.5 eV each. The Mn spectra is broader compared to Ni due to greater multiplet splitting.
The experimental spectra is well reproduced for ${\Delta}$=2.5 eV and $V$=2.1 eV along with $U_{dd}$ = 6.5 and and $U_{dc}$ = 8.5 eV. Moroever the intensity of the satellite indicates that the system can be described by a pure Mn$^{4+}$ configuration. Based on the values of the above four parameters, the main peak can be attributed to majority $cd^{4}L$ character, while the satellite peak can be attributed to $cd^{3}L$ characters. Additional features indicating existence of Mn$^{3+}$ ions is not so clearly seen in our XPS spectra.
This is unlike the case of doped rare earth manganite systems where the satellite feature of the Mn $2p$ spectra can be expressed as a linear combination of Mn in 3+ and 4+ valence states\cite{Abbate2002}.
 The weights of the $d^{3}$, $d^4L$ configurations in the ground state are 0.625 0.375 respectively, which yields the total electron number $<$$n_{d}$$>$ = 3.38 in the ground state. Thus the higher $d$ electron count than the ionic value of 3, arises due to the greater charge transfer character and hybridization in the Mn-O bonds. The relatively smaller value of the charge transfer energy is comparable to the values obtained in isostructural Mn$^{4+}$ systems viz. CaMnO$_{3}$ and SrMnO$_{3}$ \cite{Park1996}. Thus the MnO$_{6}$ octahedra possesses a greater covalency character as compared to NiO$_{6}$ in Pr$_{2}$MnNiO$_{6}$.
\subsubsection{Mn $2p$ XAS spectra}
\begin{figure}[!h]
\begin{center}
   	\includegraphics[height=0.6\textwidth]{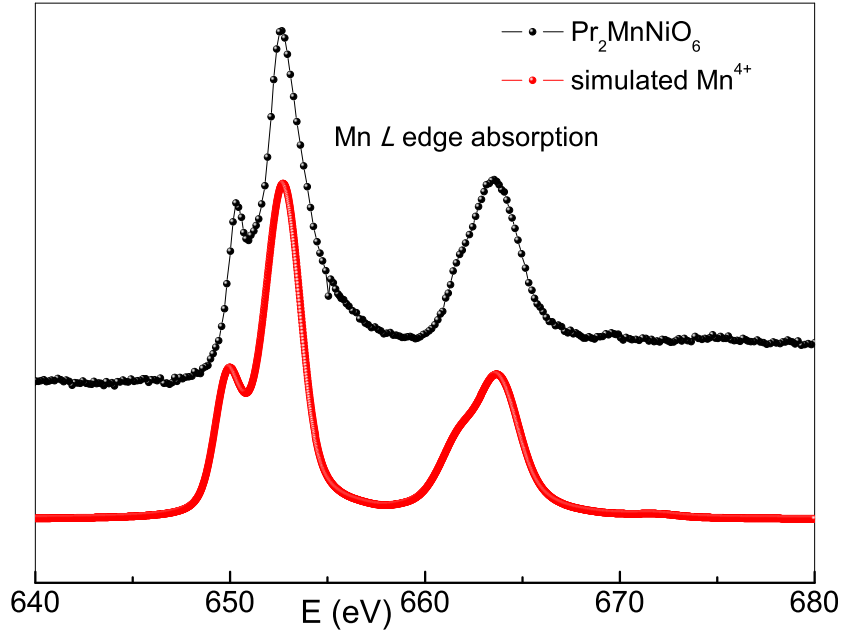}
	\caption{\textit{Mn $L_{3,2}$ absorption spectra of Pr$_{2}$NiMnO$_{6}$ along with simulated spectra. }}
\end{center}
\end{figure}
The Mn $2p$ absorption spectra of Pr$_{2}$MnNiO$_{6}$ is shown in fig. 7. The spectra comprises of two main features corresponding to $2p_{3/2}$ and $2p_{1/2}$. Unlike the XPS spectra, the absorption spectra does not show any satellite features. 
%For comparision, we have also shown the absorption spectra of parent manganite LaMnO$_{3}$.
%
The spectral feature is similar to that observed in CaMnO$_{3}$, without any distinct sign of Mn$^{3+}$ features\cite{deGroot1992}.
The spectra does not show any distinct satellite features due to charge transfer effect.
The Mn $2p$ XAS spectrum is also theoretically simulated for $d^{3}$ and $d^{4}L$ configurations in the ground state, similar to the XPS spectra. 
The Mn $2p$ absorption spectra is highly sensitive to the crystal field splitting $10D_{q}$, unlike the XPS spectra. The feature at 650 eV becomes prominent with increase in the bare crystal field term 10$Dq$. The experimental spectra is successfully reproduced for $10D_q$ = 2.4 eV. This value is similar to that used for simulation in the case of La$_{0.9}$Ca$_{0.1}$MnO$_{3}$, which is predominantly an Mn$^{4+}$ system\cite{deGroot1992}. 

%\newsection
 \begin{table*}
\caption{Best fit parameters obtained from Ni and Mn $2p$ XPS and XAS spectra for Pr$_{2}$MnNiO$_{6}$.}
\begin{center}
\begin{tabular}{p{3.0cm}c c c c c c c c c c c c c c c c}\hline\hline 
 & \multicolumn{1}{c}{} & \multicolumn{2}{c}{} & \multicolumn{5}{c}{} & \multicolumn{2}{c}{} & \multicolumn{1}{c}{} & \multicolumn{1}{c}{}\\
Compound & Ni & Mn \\ \hline 
 %Pr$_{2}$MnNiO$_{6}$& $2p_{3/2}$ &  $2p_{1/2}$ & & $2p_{3/2}$ &  $2p_{1/2}$ & & & &  & &  \\  \hline 
  ${\Delta}$  &3.5 & 2.5     & \\
  ${\Delta}_{eff}$ & 4.5 & 1.5 \\
  $U_{dd}$  & 7.5 & 6.5 \\
  $U_{eff}$ & 5.5 & 6 \\
 $n_{d}$            &  8.21                                         &3.38  \\
   $V_{e_g}$/$V_{t_{2g}}$          & 2.1                                     &2.0  \\
   $V_{pd\sigma}$   &       1.21   &                                 1.2  \\
\hline\hline
\end{tabular}
\end{center}
\end{table*}

\subsection{O $1s$ spectra and density of states}
\begin{figure}[!h]
\centering
\begin{minipage}{.55\textwidth}
  \centering
  \includegraphics[width=1.15\linewidth]{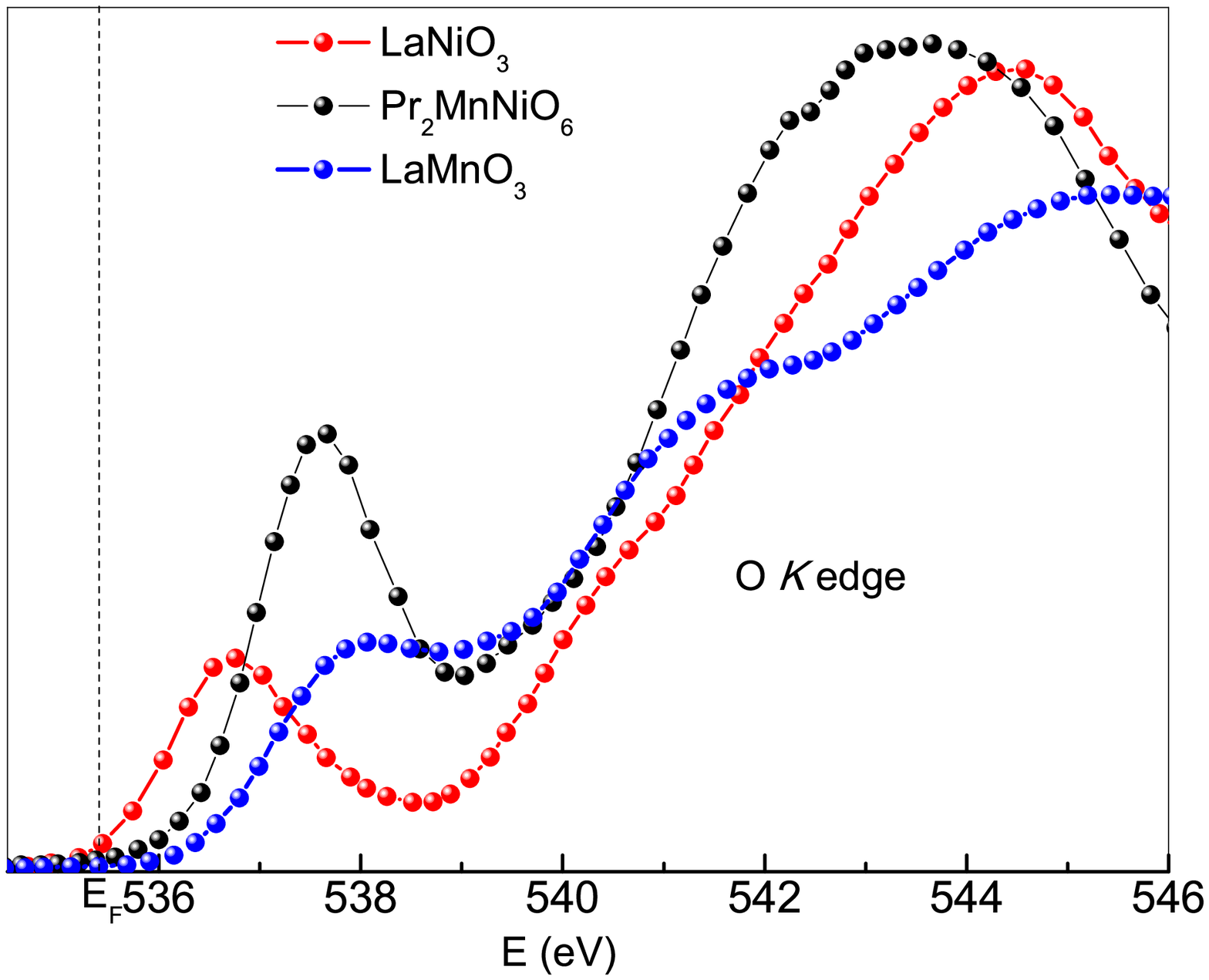}
  \label{a}
\end{minipage}%
\begin{minipage}{.55\textwidth}
  \centering
  \includegraphics[width=1.0\linewidth]{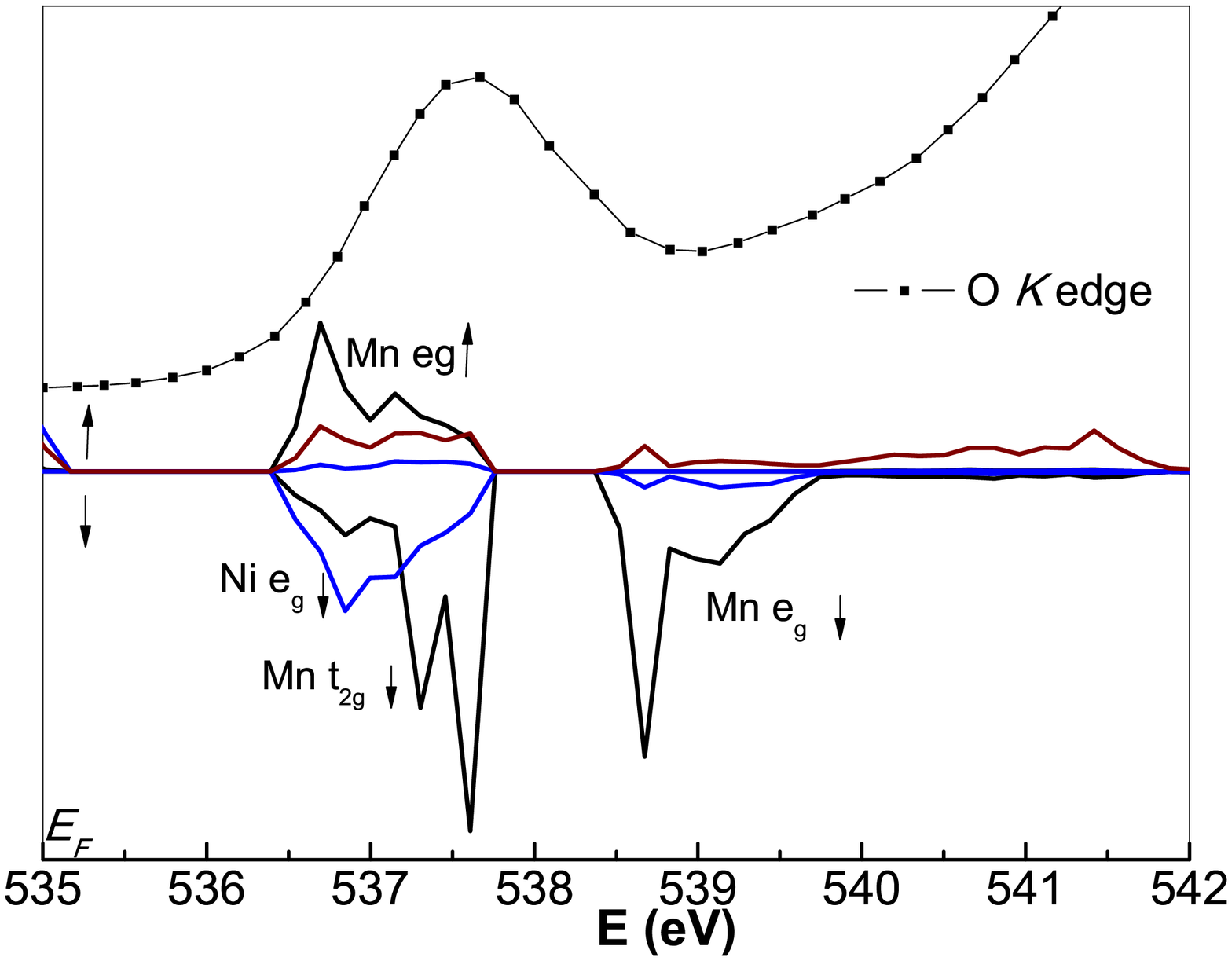}
  \label{b}
\end{minipage}
\caption{\textit{(a)The O $1s$-edge absorption spectra of Pr$_{2}$NiMnO$_{6}$, along with LaNiO$_{3}$ and LaMnO$_{3}$. The fermi energy is marked with rising edge of  metallic LaNiO$_{3}$. (b) Enlarged pre-edge region, highlighting the Mn(Ni)$3d$-O$2p$ hybridized unoccupied states and corresponding density of states above Fermi energy for $U$$-$$J$= 2 eV.}}
\end{figure}
\begin{figure}
\begin{center}
   	\includegraphics[width=0.7\textwidth]{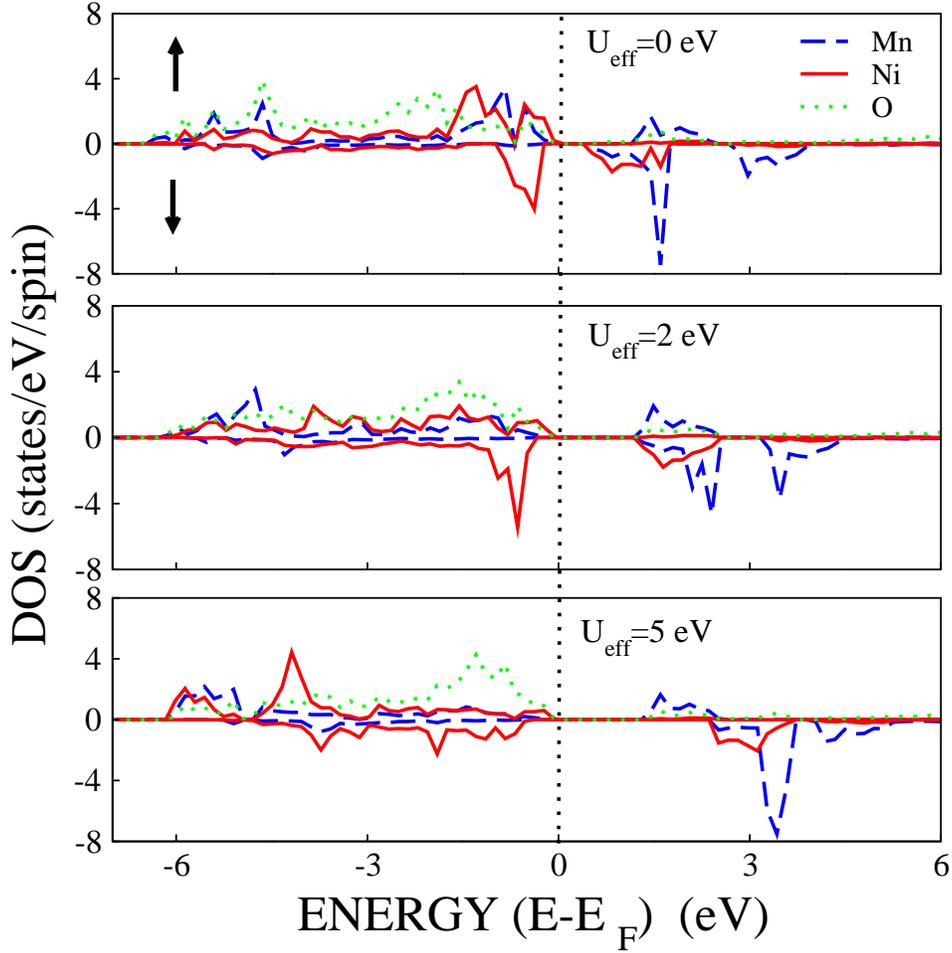}
   	\caption{\textit{ Spin polarized local density of states per formula unit of Pr$_{2}$MnNiO$_{6}$ for Mn and Ni ${\it 3d}$ states and O ${\it 2p}$ states for (a) $U_{eff}$ = 0 eV, (b) 2 eV and (c) 5 eV, where $U_{eff}$ = $U$$-$$J$.  The zero is the Fermi energy E$_{F}$.}}
\end{center}
\end{figure}
Fig. 8a shows the normalized O $1s$ edge spectra of Pr$_{2}$MnNiO$_{6}$ collected in the total electron yield mode. 
The first prominent peak of the (536-539 eV) spectra arises due to the transition from O $1s$ state to unoccupied O $2p$ states that are hybridized with $3d$ states of Mn and Ni above the Fermi energy($E_{F}$). For determination of $E_{F}$, we present normalized spectra of LaNiO$_{3}$, which is metallic compound. We also show the spectra of LaMnO$_{3}$, which has a known band gap. The position of $E_{F}$ was fixed at the rising edge of LaNiO$_{3}$ as shown in fig. 8a. The rise in spectra of LaMnO$_{3}$ occurs around 1.1 eV above $E_{F}$, which can thereby be considered as its bandgap, which is close to the bandgap of 1.2 eV obtained from optical conductivity measurements\cite{Jung1997}. For comparision of the spectra of three compounds, normalization was done at the post edge above 570 eV, which corresponds to a continuum.
  Based on the rising edge of spectra, (10${\%}$ of peak intensity) the estimated bandgap of Pr$_{2}$MnNiO$_{6}$ is nearly 0.9 eV which 
  is around 0.6 eV greater than the value obtained from resistivity measurements. However the band gap is much lower than the reported
  value of 1.4 eV in the case of La$_{2}$MnNiO$_{6}$ thin films\cite{Chungfeng2015}. The experimental band gap is also affected by presence
  of anti-site disorders, which are insulating regions. 
  
The first main peak in Pr$_{2}$MnNiO$_{6}$ occurs around 2.5 eV above $E_{F}$ as seen in fig. 8a.
The peak intensity is almost twice that of LaNiO$_{3}$ and LaMnO$_{3}$. Along with lower $d$ electron count in the Mn$^{4+}$ ion, the intensity is also affected by the larger covalency between the Mn-O and Ni-O bonds.  
The intensity of the pre-edge peak is roughly proportional to ${\beta}_{0}^2$, which is an indirect measure of the covalency of the ground state\cite{Medarde1992}. Thus large intensity of the O $1s$ pre-peak edge indicates a greater covalent character of our system as compared to LaMnO$_{3}$. This is also valid in the case of O $1s$ edge of PrMnO$_{3}$\cite{Toulemonde1999}, since in $R$MnO$_{3}$ variation in $R$ does not drastically affect the covalency character and the bandgap.  
The major contribution to the spectral intensity arises from the unoccupied Mn $e_{g}$ (${\uparrow}$ and ${\downarrow}$) and $t_{2g}$ ${\downarrow}$ states while a smaller contribution arises from the Ni $e_{g}$ ${\downarrow}$ states.
Thus the large enhancement of the unoccupied states above $E_{F}$ indicating large overlap between the Mn $3d$ and Ni $3d$ bands in Pr$_{2}$MnNiO$_{6}$.

Fig. 9 shows the spin resolved partial density of states of Pr$_{2}$MnNiO$_{6}$ comprising of the $3d$ states of Mn and Ni along with O $2p$ states for three values of $U$$-$$J$ viz. 0, 2 and 5 eV. The density of states show sufficient mixing between the Mn and Ni states near $E_{F}$. The Ni and Mn $3d$ states split into t$_{2g}$ and e$_{g}$ states due to crystal field splitting.
Below $E_{F}$, the spin up channel is occupied by the Mn $t_{2g}$${\uparrow}$ along with Ni $t_{2g}$${\uparrow}$ and partially filled $e_{g}$${\uparrow}$ states.
We observe that below -2 eV the O $2p$ states contribute significantly to the density of states. 
The Ni e$_{g}$${\uparrow}$ states occur close to $E_{F}$ in the range 0 to -1 eV with a peak at ~-0.8 eV. On the other hand the Ni t$_{2g}$${\uparrow}$ shows a prominent peak at -1.2 eV. The spin down channel below E$_{F}$ is dominated by the Ni $t_{2g}$${\downarrow}$ states.
Above $E_{F}$, the Mn e$_{g}$${\uparrow}$ states dominate the spin-up channel occuring at around 1.5 eV above $E_{F}$. However the  the Mn $t_{2g}$${\downarrow}$ states from the spin-down channel are closer to $E_{F}$(+0.5 eV). This shows a strong overlap with the Ni $e_{g}$${\downarrow}$ states. The Ni states show a greater delocalization as compared to the Mn states, which affects the magnetic moment. 
From our first principles calculations, the the Mn moments show a value of 2.8${\mu_B}$ which is close to its ionic value +3. However the Ni moments show a lesser value of 1.4${\mu_B}$ compared to its ionic value of 2.0 ${\mu_B}$. 

  Even for $U$$-$$J$ = 0 eV, Pr$_{2}$MnNiO$_{6}$ shows a band gap of 0.5 eV which is similar to the values obtained in La$_{2}$MnNiO$_{6}$\cite{HenaDas2008}. This is unlike the parent manganites which show a ferromagnetic metallic ground state in the absence of correlations. However the band gap is much smaller than the isovalent CaMnO$_{3}$ or NiO due to the overlap of Ni and Mn $3d$ states. In addition, 
the effect of Coulomb correlations is more complex due to the inequivalent nature of the two transition metal ions. To probe its effect, the calculations were also performed for different values of $U$-$J$ ie 2, 5 and 8 eV.
 With increase in $U$$-$$J$,the spectral weight of O $2p$ states increase below $E_{F}$. In addition, there occurs a shift in Ni and Mn $3d$ states which become highly localized. Also for $U$$-$$J$ = 2 eV and above, our calculations show a higher band gap of 1 eV, which remains constant even at 8 eV(not shown). The constant band-gap indicates absence of any effect on the Mn $e_{g}$${\uparrow}$ states above $E_{F}$. This is unlike the parent manganites, in which show a greater increase in band gap with increase in $U$ in a systematic manner. 
Since the O $1s$ pre-edge is a direct representation of the unoccupied DOS, we compare the Mn $3d$, Ni $3d$ and O $2p$ DOS in fig. 8b for $U$$-$$J$ = 2 eV. The nature of spectra qualitatively for $U$$-$$J$ = 2 eV since for this value we obtain a band gap of approximately 0.9 eV which is comparable to the rising edge of O $1s$ edge spectra.
Based on the DOS the O $K$ pre-peak can be divided into two portions. The first part comprises of the strongly overlapping Mn $e_{g}$${\uparrow}$ and Ni $e_{g}$${\downarrow}$ states. The large rise in the central portion can be attributed entirely to the Mn $t_{2g}$${\downarrow}$ states while the subsequent edge arises due to the Mn $e_{g}$${\downarrow}$ states. However unlike in parent manganite compounds, there is no effect of $U$ on Mn $e_{g}$${\uparrow}$ states. 
\subsection{Role of charge transfer, covalency in band gap}
In Pr$_{2}$MnNiO$_{6}$, the Ni$^{2+}$ ion unlike in NiO has smaller charge transfer energy due to which the $d^{9}L$ state has significent occupancy. Similarly, the $d^{3}$ states are strongly hybridized to the $d^{4}L$ states in the Mn$^{4+}$ ion. The band gap of the material is determined by interplay of $U_{dd}$, ${\Delta}$ and $V$, the later two reflecting the degree of covalency between the Ni-O and Mn-O bonds.  
An estimate of the hybridization strength is obtained from the Slater-Koster method, from the Ni(Mn) $3d$-O $2p$ transfer integral $V_{pd\sigma}$ ${\sim}$ ${\beta}_{0}$ ${\sim}$ 1/$d_l^{3.5}$, where $d_l$ corresponds to the average Ni-O and Mn-O bond lengths. 
The average Ni-O and Mn-O bond-lengths in Pr$_{2}$MnNiO$_{6}$ which are 2.04 and 1.9{\AA} respectively, yield $V_{pd\sigma}$ = 1.13 eV for Ni-O and a much higher value of 1.8 eV for Mn-O bonds respectively. 
However the cluster analysis of $2p$ XPS spectra yield nearly equal values, ie $V_{pd\sigma}$ = 1.21 eV for both the ions. In case of Ni$^{2+}$-O bond, $V_{pd\sigma}$  is comparable to its bare Slater-Koster integral, and also close to the value obtained for NiO from cluster analysis\cite{Bocquet1992}. However the Mn$^{4+}$-O bond in our system shows considerable reduction in $V_{pd\sigma}$, when compared to values obtained for isovalent SrMnO$_{3}$ and CaMnO$_{3}$. For the Mn$^{4+}$-O bond in both these compounds, $V_{pd\sigma}$ is in the range between 1.5 and 1.6 eV, as obtained from XPS and XAS spectra\cite{Saitoh1995, Abbate2002}.

Thus CaMnO$_{3}$ and SrMnO$_{3}$ with a much smaller ${\Delta}_{eff}$ show a strongly hybridized ground state due to which they are considered as charge transfer insulators, though with a much larger bandgap. In the case of parent LaMnO$_{3}$, a large value of 2.2 eV is obtained for $V_{pd\sigma}$, but due to relatively higher value of ${\Delta}_{eff}$ the $R$MnO$_{3}$ compounds have a mixed character, ie between that of a Mott-Hubbard and charge transfer insulator in the ground state\cite{DDSarma1993}.

In the case of $R$NiO$_{3}$, the scenario is entirely different. Based on the average Ni-O bondlength of 1.94{\AA}, bare hybridization strength of 0.9 eV is obtained in the case of PrNiO$_{3}$. However the cluster calculation yields a ground state with large value of ${\beta}^2$ ${\sim}$ 0.55 and $V_{pd\sigma}$ = 1.5 eV, which is much greater than the bare Slater-Koster transfer integrals.
The large covalency due to a small transfer energy (${\Delta}$ ${\sim}$ 1 eV) and Ni-O-Ni inter-cluster hopping are more sensitive to structural variation in nickelates\cite{Neidmayer1992}. 
Increase in Ni-O bond length and reduction in Ni-O-Ni bond angles from La to Nd drastically affect the hopping integrals and the $e_{g}$ bandwidth\cite{DDSarma1994, DDSarmaJPCM1994}. Thus with varying $R$ from La to Nd, this results in a transition from metallic to insulating state with a very small band gap in PrNiO$_{3}$ and NdNiO$_{3}$.
 Due to this, PrNiO$_{3}$ and NdNiO$_{3}$ are considered as ``covalent insulators" which are intermediate between charge transfer insulator and $p$$-$$d$ metal in the ZSA diagram \cite{DDSarmaJPCM1994}.

However in the case of $R$$_{2}$MnNiO$_{6}$, the changes in Mn(Ni)-O bond lengths and Mn-O-Ni bond angles do not affect the Mn-O-Ni hopping integrals and the band width of the $e_{g}$ orbitals, even though there occurs changes in super-exchange strength, which reduces the magnetic transition temperatures.
This is also seen indirectly from high pressure studies on La$_{2}$MnNiO$_{6}$ which retains its ferromagnetic character even under 30 GPa pressure, with a small variation in $T_{C}$ and reduction in magnetic moments\cite{Haskel2011}. 
Thus in the $R$$_{2}$MnNiO$_{6}$ based double perovskites, the charge transfer from oxygen to Mn$^{4+}$ and Ni$^{2+}$ ions have a more robust character due to which the bandgap remains largely unaffected.

Based on the scheme of ZSA model, in many of the Ni$^{2+}$-based compounds(except NiO), the band gap in terms of the simple charge transfer model can be expressed as, $E_{g}$ = ${\Delta}$+${\delta}$-$W$/2; where $W$ corresponds to the ligand band-width, and ${\delta}$=2${\delta}^{n}$-${\delta}^{n-1}$-${\delta}^{n+1}$, corresponds to lowering of energy of $3d^{8}$ configuration due to hybridization\cite{Sawatzky1985}. 
In a simplified approximation, the band gap $E_{g}$ is given by ${\Delta}_{eff}$-$W$, where $W$$=$($W_{d}$$+$$W_{p}$)/2, is the average of the transition metal and ligand bandwidths\cite{Fujimori1993}. From the single impurity Anderson model, we can assume that $W_{p}$ = 4 eV and $W_{d}$ = 0.5 eV (for Ni and Mn $3d$ bandwidths).
Thus if we consider ${\Delta}_{eff}$ of Ni$^{2+}$ ion alone in Pr$_{2}$MnNiO$_{6}$, we would obtain a band gap of 2.25 eV, which though smaller than NiO by half still is much greater than the experimental value of 0.9 eV. 
Due to the highly reduced ${\Delta}_{eff}$ of Mn$^{4+}$ ion, we can consider a net charge transfer energy of the system which is average of Mn and Ni. Thus for a net average charge transfer energy, ${\Delta}_{eff}^{av}$ = 3 eV, we obtain a band band gap of 0.75 eV, which is much closer to the experimentally obtained values, thereby suggesting a Ni$^{2+}$-O-Mn$^{4+}$ charge transfer effect. 
The effect of charge transfer nature is also seen from the DOS in fig. 9. At $U$$=$0 eV, since the highest spectral weight is of $3d$, the gap is of $d-d$ type. With increase in $U$, there is a large shift in spectral weight from $3d$ to O $2p$ states. 
Thus from first princples along with XPS studies we can establish that the double perovskite Pr$_{2}$MnNiO$_{6}$ is an intermediate covalent compound according to the ZSA diagram, with a band gap which is of ``$p-d$'' type.
\subsection{Conclusions}
To summarize, the electronic structure of double perovskite compound Pr$_{2}$MnNiO$_{6}$ is studied using $2p$ core XPS and XAS along with O $K$ edge absorption. The Ni $2p$ XPS shows a three peak structure indicating a greater charge transfer effect. 
Using the charge transfer multiplet theory it is found that Ni$^{2+}$ has a lower charge transfer energy of 3.5 eV as compared to NiO compound resulting in a ground state of Ni is 78${\%}$ $d^8$, 21${\%}$ $d^9L$ and 0.6${\%}$ $d^9L^2$. Similar analysis of Mn $2p$ XPS reveal that Mn$^{4+}$ has a charge transfer energy of 2.5 eV, which is close to that of CaMnO$_{3}$ and SrMnO$_{3}$. The ground state of Mn is 62${\%}$ $d^3$ and 38${\%}$ $d^4L$.
The ground state of Ni$^{2+}$ and Mn$^{4+}$ reveal a much higher $d$ electron count of 8.21 and 3.38 respectively. 
 Based on our cluster analysis, we estimate a band gap of 0.75 eV which is close to the value obtained from the O $K$ edge spectra.
Since for Mn$^{4+}$ and Ni$^{2+}$, $U$$>$${\Delta}$ and $U_{eff}$$>$${\Delta}_{eff}$ conditions are satisfied, Pr$_{2}$MnNiO$_{6}$ can be considered as a charge transfer insulator, with an intermediate covalent character.
The density of states reveal a band gap of nearly 1 eV for $U$$-$$J$$>$2 eV, and remains constant even for $U$$-$$J$=8 eV. 
Similarly our density functional theory calculations show that the band gap is of $p$-$d$ type confirming the more robust charge transfer nature.

\end{document}